\documentclass{article}

\usepackage{arxiv}

\usepackage[utf8]{inputenc} % allow utf-8 input
\usepackage{booktabs}       % professional-quality tables
\usepackage{nicefrac}       % compact symbols for 1/2, etc.
\usepackage{microtype}      % microtypography

\usepackage{graphicx}
\usepackage{upgreek}
\usepackage{multicol,multirow}
\usepackage{amsmath,amssymb,amsfonts}
\usepackage{mathrsfs}
\usepackage{amsthm}
\usepackage[figuresright]{rotating}
\usepackage{appendix}
\usepackage[authoryear]{natbib}
\usepackage{ifpdf}
\usepackage[T1]{fontenc}
\usepackage{newtxtext}
\usepackage{newtxmath}
\usepackage{textcomp}
\usepackage{xcolor}
\usepackage{xurl}
\usepackage[colorlinks,allcolors=blue]{hyperref}
\definecolor{jourcolor}{cmyk}{1,0.57,0.01,0.38}
\hypersetup{
    colorlinks,%
    citecolor=jourcolor,%
    filecolor=jourcolor,%
    linkcolor=jourcolor,%
    urlcolor=jourcolor
}

\theoremstyle{definition}

\usepackage{siunitx}
\usepackage{lipsum}
\usepackage[nameinlink, capitalise, noabbrev]{cleveref}

\newcommand{\ti}{\mathrm{TI}}
\newcommand{\sxv}{\alpha_s \cdot \alpha_v}

\title{Predict future sale}

\author{
 Kirby S. Heck$^1$ and Michael F. Howland$^{1*}$ \\
 \\
  $^{1}$Civil and Environmental Engineering, Massachusetts Institute of Technology, Cambridge, MA 02139, USA \\
  $^*$Corresponding author: \texttt{mhowland@mit.edu} \\
}

\title{Unraveling the effects of atmospheric dynamics on wakes with a controlled synthetic inflow methodology}

\begin{document}

% \author[Kirby S. Heck and Michael F. Howland]{Kirby S. Heck$^{1}${{\href{https://orcid.org/0009-0002-8719-2967}{\includegraphics{orcid_logo}}}} and Michael F. Howland$^{1\ast}${{\href{https://orcid.org/0000-0002-2878-3874}{\includegraphics{orcid_logo}}}}}

% \address[1]{Civil and Environmental Engineering, Massachusetts Institute of Technology, Cambridge, MA 02139, USA}

% \corres{*}{Corresponding author. E-mail:
% \emaillink{mhowland@mit.edu}}

% \keywords{}

% \date{\textbf{Received:} XX 2020; \textbf{Revised:} XX XX 2020; \textbf{Accepted:} XX XX 2020}

\maketitle

\begin{abstract}
Winds in the atmospheric boundary layer (ABL) display a wide range of velocity profiles and turbulence properties that affect wind turbine wake dynamics.
However, standard concurrent-precursor large eddy simulations (LES) often neglect phenomena such as mesoscale patterns, limiting the range and controllability of inflow parameters that can be studied. 
Here, we propose a synthetic inflow LES method with high inflow controllability to allow parameters such as shear, turbulence, and Coriolis effects to be varied independently, facilitating the efficient exploration of wake dynamics across the full range of conditions observed in the field. 
The synthetic inflow method faithfully reconstructs wake dynamics when compared with standard concurrent-precursor LES. 
We then run a suite of over 600 LES cases to investigate the ABL processes that most affect wake dynamics. 
We find that wake recovery strongly depends on inflow wind veer, especially at low turbulence intensities, due to the elongation of the skewed wake. 
Furthermore, we identify a novel scaling relation that collapses wake deflections and dynamics onto the combination of shear and veer. 
The suite of LES cases elucidates ABL regimes and wake dynamics where current and future wind turbines may operate, building toward improved wake modeling for wind farm design and control. 
\end{abstract}

\vspace{8pt}
% \begin{boxtext}
\textbf{\mathversion{bold}Impact Statement}
Atmospheric winds vary drastically in speed, direction, and turbulence properties, significantly affecting wind turbine wakes---regions of slower-moving air behind the rotor. 
However, existing simulation approaches struggle to span the full complexity of these atmospheric conditions. 
We present a new method for simulating wake dynamics by prescribing a wind profile and layering turbulence and additional physics, such as Coriolis forces, giving full control over the incident wind. 
By analyzing over 600 simulations, we isolate critical atmospheric processes that most strongly affect wake evolution. 
These insights provide a scalable framework for understanding wake physics across the wide range of conditions present in the atmosphere, which may lead to improved wake modeling and improved wind farm efficiency.

\section{Introduction}
% - Introduce the field.
	% - Introduce the problem.
	% - What is the knowledge gap?
	% - What are you doing about it?
	% - How are you doing it?
	% - Why is it important?

Wind turbines, which extract energy and momentum from winds in the atmospheric boundary layer (ABL), cast wakes downstream. 
These energy-deficient regions negatively impact power production downwind, while elevated levels of turbulence may increase fatigue loading. 
Mitigating wake interactions within wind turbine arrays relies on understanding how energy and momentum are replenished into the wake region through turbulent mixing. 
However, wake transport mechanisms are affected by a large set of ABL processes, including wind shear and wind veer (vertical gradients of wind speed and wind direction, respectively), stratification, Coriolis forcing, mesoscale forcing, orography, and diurnal and transient effects, to name a few \citep{landberg_meteorology_2015}. 
These properties of the ABL are often correlated, posing a challenge for studying each effect in isolation \citep[c.f.,][]{turk_dependence_2010, kelly_shear_2023}. 
Additionally, mesoscale forcing and diurnal effects contribute to a high degree of variability for the inflow properties of the ABL, which are difficult to simulate with microscale large eddy simulations (LES) or recreate in wind tunnel experiments \citep{haupt_lessons_2023}. 
As a result, the effects of individual, measurable properties of the ABL, such as wind veer, on the resulting wake dynamics have not been systematically identified in controlled tests. 

Many previous studies have investigated ABL--wake interactions, most commonly through the concurrent-precursor method \citep{stevens_concurrent_2014}. 
In the concurrent-precursor method, an empty domain is used to spin up a turbulent ABL inflow, which serves as the inflow condition to a second domain containing one or more wind turbines. 
A key focus of previous work has been the effects of turbulence intensity ($\ti$) and wind shear on wake recovery in the ABL. 
For example, \citet{wu_atmospheric_2012} used LES to investigate the effect of shear and $\ti$ by varying the roughness length $z_0$ in neutral boundary layers, finding that across the range of $\ti$ and wind shear studied, $\ti$ has a strong effect on wake recovery, while the effect of wind shear on wake recovery is comparatively negligible. 
These conclusions are corroborated by other numerical studies \citep{churchfield_numerical_2012, chu_turbulence_2014} and wind tunnel studies \citep{chamorro_wind-tunnel_2009, bartl_wind_2018}. 
Several LES studies have investigated the effects of stratification on wake development \citep{churchfield_numerical_2012, abkar_influence_2015}, finding that under convective conditions, vertical meandering and wake recovery are enhanced, while under stably stratified conditions, wake spreading and recovery are suppressed. 
However, wind shear, inflow $\ti$, and turbulence length scales are simultaneously affected when varying the background stratification.
To disentangle these effects, \citet{du_influence_2021} strategically varied $z_0$ and the domain height $L_z$ in concurrent-precursor LES to match the hub height wind speed $U_h$ and $\ti$ across stable, neutral, and unstable (convective) conditions. 
As a result, they concluded that larger turbulence length scales were responsible for faster wake recovery in convective conditions, relative to neutral conditions, despite lower shear, and vice-versa in stable conditions. 
Similarly, increasing turbulence length scales were found to increase wake recovery in wind tunnel experiments and LES by \citet{espana_spatial_2011} and \citet{vahidi_influence_2024}, respectively, and to promote wake meandering in wind tunnel experiments \citep{bourhis_impact_2025}. 

Efforts to perform controlled experiments of ABL inflow properties are further complicated by the introduction of Coriolis forces. 
\citet{xie_numerical_2017} investigated the effects of stratification across unstable, neutral, and stable conditions with the inclusion of Coriolis forces in a five-turbine wind farm. 
They found similar power losses due to wake effects between neutral and stable conditions, but decreased power losses due to enhanced wake recovery in convective conditions. 
However, they noted many differences in inflow properties resulting from varying the stratification, including hub height velocity, $\ti$, length scales, and wind shear and veer. 
Other LES studies have observed that introducing Coriolis forces results in enhanced wake recovery, relative to an inflow without Coriolis forces, due to the presence of wind veer \citep{abkar_influence_2016, churchfield_effects_2018}. 
\citet{klemmer_momentum_2024} found that in stably stratified ABLs, wind shear and veer significantly alter turbulence and momentum budgets in the wake, enhancing the additional turbulence generated by the turbine wake relative to neutral ABLs. 
Several studies have also noted that the indirect effects of stability on the inflow, such as through wind shear, veer, and turbulence length scales, have a more pronounced effect on the wake dynamics than the direct buoyant production/suppression of turbulence kinetic energy (TKE) \citep{du_influence_2021, klemmer_momentum_2024, klemmer_wake_2025}. 
In neutral ABLs, \citet{heck_coriolis_2025} found that changing the relative strength of Coriolis forcing had a minimal impact on the wake recovery rate, but significantly affected the magnitude and direction of ABL inflow-induced wake deflections depending on incident conditions. 
Other studies have also observed competing effects of Coriolis forces on wake deflections \citep[c.f.][]{van_der_laan_why_2017, howland_influence_2018, gadde_effect_2019}.
Clearly, the properties of the ABL inflow can have a substantial influence on wind turbine wake dynamics, but studying the effects of each parameter in isolation poses many challenges due to the coupled nature of ABL processes. 
Meanwhile, the wide range of stratification, surface morphology (roughnesses), and large-scale weather conditions that are present in real ABLs causes a significant spread in these inflow parameters that cannot all be reliably produced by a microscale LES model in isolation.

Mesoscale-to-microscale coupled (MMC) simulations introduce transient dynamics and large-scale weather patterns into the LES flow, disrupting the correlated nature of flow variables in microscale simulations, such as wind shear and wind veer \citep{landberg_meteorology_2015}. 
As a result, a wider variety of wind conditions can be simulated in MMC simulations compared with microscale LES alone. 
For example, \citet{wise_meso-_2022} used MMC simulations to investigate wind turbine wakes over a mountain ridge in stable and convective conditions. 
The authors found that in stable conditions, wakes follow the terrain contours, but in convective conditions, wakes are carried aloft. 
Recently, \citet{chatterjee_modeling_2025} used coupled simulations to investigate the impact of low-level jets on wind farm performance. 
While MMC simulations are useful for case studies and faithful reconstruction of atmospheric dynamics, they have limitations. 
In addition to the elevated computational cost of performing nested domain simulations, synthetic turbulence perturbation methods are critical to reconstructing the spectral content in the microscale domain \citep{munoz-esparza_bridging_2014, munoz-esparza_stochastic_2015}. 
Furthermore, the added complexity of the multiscale coupling can result in a loss of controllability for simulating a target inflow profile, and the transient nature of MMC simulations limits the usefulness of turbulence statistics and budgets.

Synthetic inflow generation is another alternative to studying ABL--wake interactions across inflow conditions. 
Often, a mean wind speed profile is prescribed with turbulent perturbations superimposed, and a region of the streamwise domain is allotted to flow adjustment for the turbulence to equilibrate \citep{troldborg_simple_2014}. 
The injected turbulence may derive from a precursor simulation, from forced homogeneous, isotropic turbulence (HIT), or from synthetic turbulence generation methods, such as the Mann method \citep{mann_wind_1998} or the divergence-free synthetic eddy method \citep{poletto_new_2013}.
Several studies have used synthetic turbulence generation to study wake dynamics separate from the ABL \citep[e.g.,][]{keck_synthetic_2014, ghate_interaction_2018, doosttalab_interaction_2020, li_onset_2022, li_resolvent-based_2024}. 
For example, \citet{troldborg_simple_2014} found excellent agreement between field measurements of a wind turbine wake taken by meteorological masts and a synthetic ABL inflow using the Mann turbulence method. 
However, no studies have extended the synthetic ABL methodology to include additional properties of the inflow beyond wind speed shear. 

In this study, we propose a synthetic inflow approach to studying ABL--wake physics across broad ranges of wind shear, wind veer, Coriolis forcing, and turbulence parameters. 
In prescribing a mean wind speed profile, we decouple the inflow shear from the effects of the wall (prescribed here as a slip-wall) and therefore the mechanism for turbulence generation. 
Instead, turbulence is injected into the domain at the inlet with a prescribed integral length scale and $\ti$. 
The wind direction profile is also prescribed, decoupling the wind veer from the combined effects of stratification, Coriolis forcing, and wind shear. 
We first verify that the synthetic inflow approach reproduces the wake structure and dynamics present in concurrent-precursor simulations of wind turbine wakes immersed in stratified ABLs. 
Using this methodology, we can, for the first time, explore the impact on wake dynamics from shear, veer, and other ABL parameters independently. 

The remainder of the article is organized as follows. In \cref{sec:methods}, we introduce the synthetic inflow methodology and the LES numerical setup. 
We define a wake momentum budget that is used both for method verification and for flow physics analysis. 
To guide the regimes investigated in the LES cases, we investigate range of ABL parameters observed at a field site on the north Atlantic coast of the United States. 
Beginning in \cref{ssec:results_comparison_TNBL}, we verify the synthetic inflow method against simulations of wakes immersed in ABLs simulated with a concurrent-precursor setup. 
Then, in \cref{sec:results_exploration}, we use the new LES methodology to explore a wide range of parameters to identify key variables that have the greatest impact on wind turbine wake evolution. 
Finally, conclusions and opportunities for future work are summarized in \cref{sec:conclusions}.

\section{Methodology}
\label{sec:methods}

In this section, we first introduce the LES numerical setup and the synthetic inflow methodology in \cref{ssec:methods_les}. 
The momentum budget analysis, which is used to validate the synthetic turbulence inflow method and also to analyze the evolution of wakes as a function of inflow parameters, is described in \cref{ssec:methods_budgets}. 
Then, we outline a parameterization of inflow wind shear and wind veer in \cref{ssec:methods_abl_parameterization}. 
Finally, in \cref{ssec:methods_mvco}, we describe the field data used to motivate the range of inflow parameters that are studied using the synthetic inflow method in this work. 

\subsection{Large eddy simulation setup}
\label{ssec:methods_les}

The open source, pseudo-spectral LES code Pad{\'e}Ops (\url{https://github.com/Howland-Lab/PadeOps}) is used to solve the incompressible, filtered Navier--Stokes equations with the Boussinesq approximation for buoyancy using a coupled prognostic equation for potential temperature \citep{ghate_subfilter-scale_2017, heck_coriolis_2025}.
Fourier collocation is used in the horizontal directions, and a fringe region \citep{nordstrom_fringe_1999} is used in the streamwise direction to allow for finite-length wakes and to enforce inflow boundary conditions. 
A sixth-order-accurate Pad{\'e} scheme is used for vertical differencing \citep{lele_compact_1992}, and the bottom and top boundary conditions are slip walls with a no-flux condition. 
Time integration uses a fourth-order Runge--Kutta scheme. 

\begin{figure}[htb]
    \centering
    \includegraphics[width=0.7\linewidth]{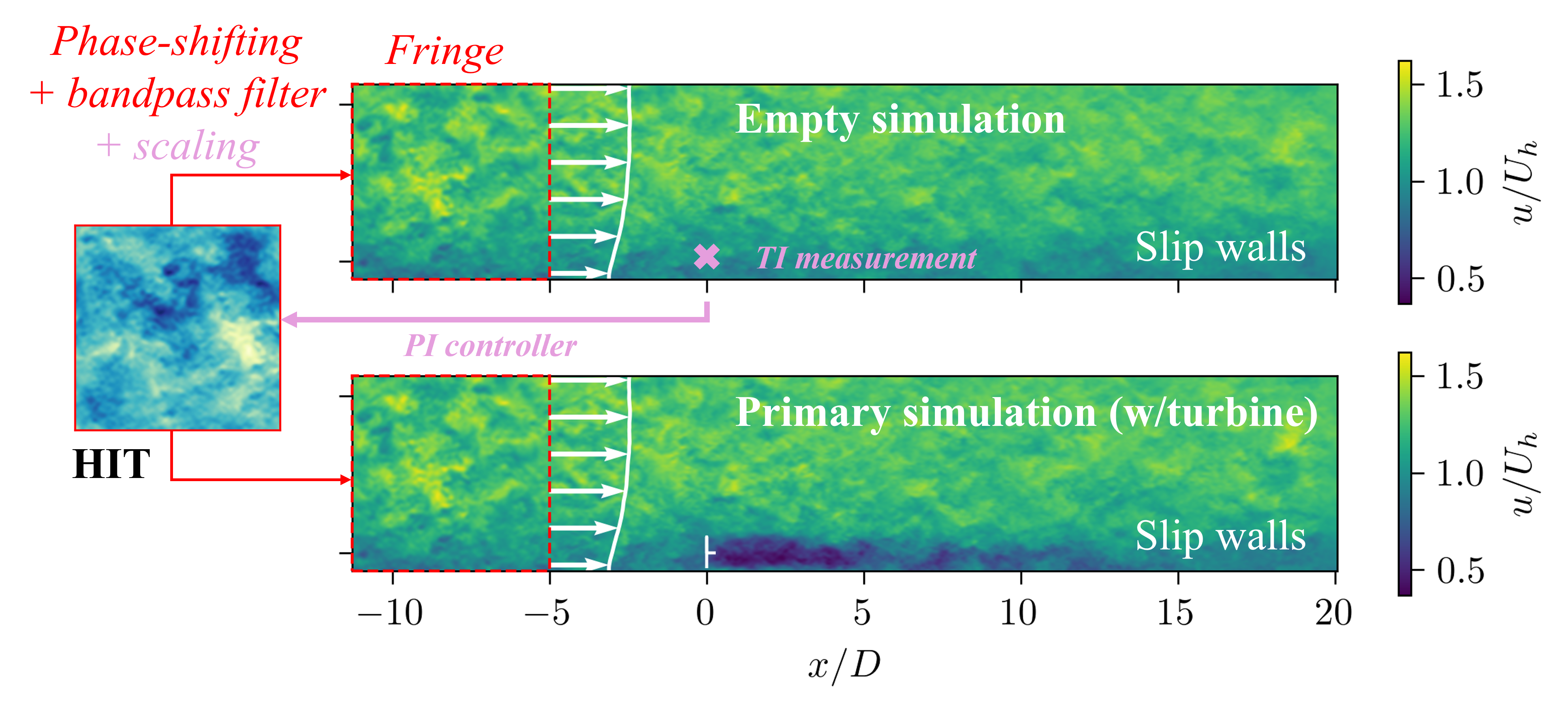}
    \caption{LES methodology for the synthetic inflow simulations. Three large eddy simulations are concurrently run (HIT, empty, primary), and the velocity perturbations from the HIT simulation are advected and superimposed on the mean profiles in the primary and empty domains.}
    \label{fig:les_setup}
\end{figure}

To investigate a wide range of combinations of turbulence intensity, wind shear, wind veer, and Coriolis forcing in the ABL, idealized, synthetic inflow profiles are generated to approximate ABL inflows. 
The mean wind speed and direction profiles in the idealized simulations are prescribed by the methodology outlined in \cref{ssec:methods_abl_parameterization}.
The geostrophic pressure gradient is varied as a function of $z$ to maintain consistency with the imposed velocity profile, thereby mitigating inflow adjustment. 
As the top and bottom boundaries are slip walls, turbulence must be injected into the domain, which is forced through a concurrently run simulation of forced homogeneous, isotropic turbulence (HIT). 
Turbulence is introduced by superimposing the prescribed mean wind profiles with the HIT wind fields, which are phase-shifted to advect the turbulent fluctuations by the mean velocity as a function of $z$. 
Furthermore, the turbulent fluctuations are bandpass-filtered by a prescribed high-pass filter of wavelength $\lambda_\mathrm{BP}$ to adjust the integral length scale \citep{ghate_interaction_2018}.
A schematic for the modified concurrent LES setup is shown in \cref{fig:les_setup}. 

Turbulence will evolve as it is advected downstream \citep{keck_synthetic_2014}. 
For example, the turbulence intensity prescribed at the inlet of the simulation will decay away with downstream distance under uniform inflow conditions \citep[c.f.][]{comte-bellot_use_1966}, but it could increase with downstream distance if the inflow speed or direction shear is sufficiently high, driven by shear production. 
To achieve a specified turbulence intensity at the rotor disk, a closed-loop proportional-integral (PI) controller is used that measures the turbulence at the location of the turbine in a concurrently run empty domain simulation with no turbine. 
The closed-loop PI controller is robust across inflow conditions and is switched off once the desired turbulence intensity at the location of the actuator disk becomes quasi-steady, as measured in the empty domain simulation. 
That is, the PI controller is turned off before any statistics are computed. 
Finally, the empty simulation is used to compute on-the-fly deficit statistics \citep{klemmer_momentum_2024}, further described in \cref{ssec:methods_budgets}. 

For the primary and empty simulations, the domain size is $10\pi D \times 4\pi D \times 2\pi D$, and contains $480 \times 192 \times 96$ points in the streamwise ($x$), lateral ($y$), and vertical ($z$) directions, respectively.
This results in an isotropic grid with a grid size of $\Delta \approx 0.065D$. 
The rotor diameter $D$ is arbitrary because there is no other length scale in the simulation setup (e.g., surface roughness; ABL height).
A transient spin-up time of 16 flow-through times is used in all simulations to allow wakes to develop and for the TI controller to reach equilibrium. 
After the initial spin-up, flow statistics are time-averaged for 24 flow-through times for wake statistics to converge. 

An irrotational actuator disk model (ADM) is used in all simulations to create a wake. 
The ADM is placed $x = 5D$ downstream of the domain inlet to allow the flow to develop from the inflow condition, and to accommodate the induction zone of the wind turbine. 
The turbine is centered laterally in the domain, and the hub height is $z_h/D = 0.625$, which corresponds to $\SI{150}{\meter}$ for the IEA~15~MW reference turbine \citep{gaertner_definition_2020}.
The magnitude of the thrust force of the turbine $F_T = \tfrac 12 \rho \pi R^2 u_d^2 C_T'$ is computed based on the disk-averaged velocity $u_d$, where $R = D/2$ is the turbine radius and $C_T' = 4/3$ is the local thrust coefficient, corresponding to $C_T = 0.75$ \citep{calaf_large_2010} in shear-free, uniform inflow conditions. 
The correction factor introduced by \citep{shapiro_filtered_2019} is used in all simulations, except when comparing to the TNBL cases from \citet{heck_coriolis_2025}, because they did not use the Shapiro correction factor in their simulations. 

\subsection{Momentum budget and streamtube analysis}
\label{ssec:methods_budgets}

In this study, we will describe the dynamics of the wake using a prognostic equation for the wake velocity deficit. 
This double decomposition, originally introduced by \citet{martinez-tossas_curled_2021}, was expanded in \citet{klemmer_momentum_2024} to include the wake-added turbulence kinetic energy budget. 
An abbreviated derivation is given here with further discussion in \citet{klemmer_momentum_2024}. 

Beginning with the filtered, incompressible Navier--Stokes equations, we introduce the wake decomposition $u_i = u_i^B + \Delta u_i$ to parse the wake into the base flow (denoted with a superscript $B$) and the deficit (wake) flow (denoted with $\Delta$). 
Then, introducing the Reynolds decomposition, both the base flow and the deficit flow are further split into time-mean and fluctuating components:
\vspace{-2pt}
\begin{align}
    u_i^B &= \overline{u_i^B} + {u_i^B}', \\
    \Delta u_i &= \overline{\Delta u_i} + \Delta u_i'. 
\end{align}
We then substitute the double decomposition into the filtered, incompressible Navier--Stokes equations, Reynolds average, and assume statistical stationarity to remove the temporal derivatives, resulting in the full RANS equations. 
Then, because the base flow must also satisfy the RANS equation, we subtract off the base flow RANS equations from the full RANS to arrive at an equation for wake velocity deficit: 
\vspace{-2pt}
\begin{align}
\newcommand*{\vp}{\vphantom{\frac{\overline{u_i'}}{x_j}}}
\label{eq:RANS_deficit}
    \underbrace{\vp (\overline{u_i^B} + \overline{\Delta u_i}) \frac{\partial \overline{\Delta u_i}}{\partial x_j}}_\mathrm{Advection} 
    = 
    \underbrace{\vp -\frac{\partial \overline{\Delta p}}{\partial x_i}}_\mathrm{Pressure~grad.}
    + \underbrace{ \vp \frac{\delta_{i3}}{\theta_0 Fr^2} \overline{\Delta \theta}}_\mathrm{Buoyancy}
    \underbrace{\vp - \frac{1}{Ro}\varepsilon_{ij3} \overline{\Delta u_j}}_\mathrm{Coriolis}
    \underbrace{\vp - \frac{\partial \overline{\Delta \tau_{ij}}}{\partial x_j}}_\mathrm{SGS}
    \underbrace{\vp - \overline{\Delta u_j}\frac{\partial \overline{u_i^B}}{\partial x_j}}_\mathrm{Base~advection}
    \underbrace{\vp - \frac{\partial \Delta \overline{u_i' u_j'}}{\partial x_j}}_\mathrm{Reynolds~stresses}.
\end{align}
Here $\Delta p$ is the wake-induced pressure field, $\Delta \theta$ is a wake-induced perturbation to the potential temperature, $\Delta \tau_{ij}$ represents the wake-added subgrid stresses (SGS), and $\Delta \overline{u_i' u_j'} \equiv \overline{u_i' u_j'} - \overline{{u_i^B}' {u_j^B}'}$ is the wake-added Reynolds stress (RS) tensor \citep{klemmer_momentum_2024}. 
Note that in \cref{eq:RANS_deficit}, only the vertical component of Earth's rotation is included \citep{howland_influence_2020}, and the Rossby number is defined $Ro = U_h/(f_c D)$, where $U_h$ is the hub-height velocity, and $f_c = 2\omega \sin(\phi)$ is the Coriolis parameter, where $\phi$ is the latitude and $\omega = \SI{7.29e-5}{\radian\per\second}$ is the rotation rate of Earth. 
The Froude number $Fr = U_h/\sqrt{gD}$ is used in the stratified simulations in \cref{ssec:results_comparison_SBL}, where $g = \SI{9.81}{\meter\per\second\squared}$ is the gravitational acceleration. 

To facilitate the analysis of the wake physics, a streamtube \citep{meyers_flow_2013} is used to spatially aggregate wake velocities, wake momentum budget terms in \cref{eq:RANS_deficit}, and identify the wake centroid location \citep{heck_coriolis_2025}. 
Tracking a streamtube selects a three-dimensional region that represents the wake robustly even in cases of inflow wind shear and wind veer. 
Throughout this study, we seed the streamtube at a radius $R_s = 0.7R$ in the rotor plane to avoid any Gaussian smoothing at the edges of the ADM \citep{shapiro_modelling_2018, heck_modelling_2023}.
Streamtube-averaged quantities are denoted by $\langle \cdot \rangle$, and are averaged within the streamtube region in the lateral ($y$) and vertical ($z$) directions.

\subsection{Atmospheric boundary layer representation}
\label{ssec:methods_abl_parameterization}

The time-mean inflow profiles are prescribed as inputs to the simulation methodology introduced in \cref{ssec:methods_les}. 
In this study, inflow profiles of streamwise ($u$) and lateral ($v$) velocity depend only on the vertical coordinate $z$, and the vertical inflow velocity ($w$) is zero. 
We consider two sources of inputs for the mean inflow profiles. 
First, to verify the synthetic inflow method against concurrent-precursor simulations of wakes in the ABL, we prescribe the synthetic inflow with the exact mean profiles from the ABL precursor simulations. 
In this setup, the primary remaining difference in the turbulent inflow of the synthetic inflow LES and the concurrent-precursor LES is the structure of the turbulence (length scales, anisotropy, etc.). 
Second, to investigate a wide range of inflow profiles, we prescribe an idealized parameterization of wind speed and wind direction profiles. 
The parameterized inflow enables efficient simulation of wind turbine wakes subjected to independently varying wind shear and wind veer strengths.

The idealized profiles parameterize the mean wind speed $U = (\bar{u}^2 + \bar{v}^2)^{1/2}$ and mean wind direction $\psi = \arctan(\bar{v}/\bar{u})$ with the following expressions: 
\begin{subequations}
    \label{eq:U_phi}
    \begin{align}   \refstepcounter{equation}
        U(z)/U_h &= 1 + (1-\epsilon) \tanh \left(\frac{\alpha_s(z - z_{h})}{D(1-\epsilon)}\right);     &
        \psi(z) &= (1-\epsilon) \tanh \left( -\frac{\alpha_v(z - z_{h})}{D(1-\epsilon)}\right).     \tag{\theequation, b}
    \end{align}     \refstepcounter{equation}
\end{subequations}
Here, $U_h$ is the wind speed at hub height $z_h$. 
The variables $\alpha_s$ and $\alpha_v$ control the vertical gradients of wind speed (shear) and wind direction (veer), respectively. 
Near $z = z_h$, $\tanh(x) \approx x$ and thus the profiles are approximately linear across the rotor extent: $U(z)/U_h \approx 1 + \alpha_s (z-z_h)/D$ and $\psi(z) \approx -\alpha_v (z-z_h)/D$. 
Note that $\alpha_z > 0$ corresponds with positive wind veer, where $\psi$ decreases as a function of height in the northern hemisphere following the Ekman spiral. 
Away from the rotor, the wind speed and direction smoothly transition to a constant value set by a regularization parameter $\epsilon = 0.2$ to prevent reverse-flow in the LES. 
Empirically, wind profiles are often approximately linear in the ABL \citep{debnath_characterization_2023, kelly_shear_2023}.
Choosing approximately linear profiles for the wind speed is useful because in the ABL, vertical profiles of velocity may be quite complex, departing significantly from canonical profile shapes (e.g., power law) due to diurnal effects, terrain, mesoscale forcing, etc. \citep{howland_conditions_2020}.
An effective parameterization of non-logarithmic or power law profiles (e.g., low-level jets) that depends on only one or two parameters could be utilized in future work with the same LES methodology.

\subsection{Range of atmospheric profile parameters and LES inputs}
\label{ssec:methods_mvco}

\begin{figure}[htb]
    \centering
    \includegraphics[width=0.9\linewidth]{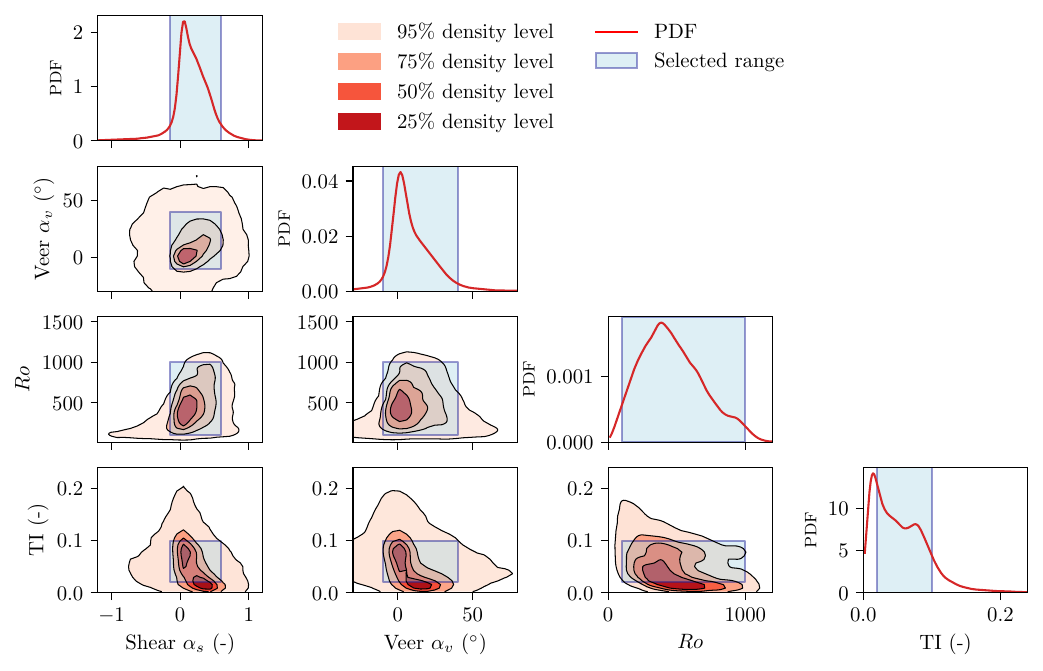}
    \caption{Joint distributions of observed ABL parameters at the Martha's Vineyard Coastal Observatory (MVCO) and profiling lidar. Shear and veer distributions are approximated as linear profiles of wind speed and direction computed across the rotor extent, while $\ti$ is interpolated at hub height. The Rossby number is based on the turbine diameter of the IEA~15~MW reference turbine.}
    \label{fig:mvco_kde}
\end{figure}

In this study, we explore wake dynamics under realistic distributions of wind shear and veer observed in field measurements. 
Field data from a profiling lidar at the Martha's Vineyard Coastal Observatory (MVCO) is used to assess distributions of ABL parameters \citep{bodini_offshore_2020, lundquist_wfip3_2023, klemmer_evaluation_2024}. 
We use data from 2017, but the joint distributions do not change significantly for the years between 2017 and 2020 (not shown). 
In this study, we chose four non-dimensional parameters relevant to modern offshore wind turbines, based on previous literature outlined in the introduction regarding the key ABL parameters that affect wake dynamics, using the dimensions of the IEA~15~MW wind turbine for reference \citep{gaertner_definition_2020}. 
The parameters of choice are the Rossby number $Ro = U_h/(f_c D)$, where $f_c$ is the Coriolis parameter, the hub-height turbulence intensity ($\ti \equiv \sqrt{2k/3}/U_h$, where $k$ is the turbulence kinetic energy), rotor-averaged wind shear $\alpha_s$, and rotor-averaged wind veer $\alpha_v$. 
Note that definitions of $\alpha_s$ and $\alpha_v$ are fit using the parameterization in \cref{eq:U_phi}($a$, $b$), respectively, where wind direction profiles are unwrapped from the original circular domain of $[0^\circ, 360^\circ)$.
These parameters are computed from the field data, which is located at a latitude $\phi = 41.4^\circ$~N. 
Lidar turbulence intensity \citep{bodini_offshore_2020} is interpolated to hub height, as is the wind speed $U_h$. 
The resulting distributions of non-dimensional inflow variables are shown in \cref{fig:mvco_kde}.

An IEA~15~MW reference turbine located at the field data site would operate, on average, in slightly sheared (0.16) and veered ($\SI{8.8}{\degree}$) conditions at relatively low TI ($5.6\%$) and a Rossby number of $\approx 470$.
However, high magnitudes of wind shear (over 0.5, i.e., 50\% variation in wind speed across the rotor) and wind veer (over $30^\circ$ across the rotor) are also common. 
Furthermore, the joint distribution of variables shows a wide spread in combinations of all variables, including shear and veer. 
The shaded regions indicate the ranges of ABL parameters chosen to simulate using LES.
Specific values of $\alpha_s$, $\alpha_v$, $Ro$, and $\ti$, as well as the turbulence length scale, which we control with the high-pass filter scale $\lambda_\text{BP}/D$ (not estimated with the lidar profiles in \cref{fig:mvco_kde}) are given in \cref{tab:LES_inputs}.
We choose a wide range of reasonable offshore turbulence intensities to simulate, recognizing uncertainties associated with inferring TKE from lidar measurements \citep{bonin_evaluation_2017}. 

\begin{table}[htb]
    \caption{Parameters defining four full factorial LES sweeps (i.e., all combinations are simulated) used in the exploration of ABL effects on wake evolution in \cref{sec:results_exploration}. Variable increments are evenly spaced (denoted with ``$\ldots$'') unless specific values are listed.}
    \label{tab:LES_inputs}
    \def\arraystretch{1.5}%  1 is the default, change whatever you need
    \centering
    \footnotesize
    \begin{tabular}{lllllll}
    \hline
    Sweep & Shear $\alpha_s$ (-)       & Veer $\alpha_v$ ($^\circ$) & $Ro$                      & $\ti$ (\%)      & $\lambda_\text{BP}/D$ & Cases \\ \hline
    A     & $-0.15, 0, \ldots, 0.6$    & $-10, 0, ..., 40$          & $125, 400, 1000$          & $3, 8$          & $2\pi/3$              & 216   \\
    B     & $0$                        & $0, 10, 20, 30$            & $1000$                    & $2, 4, ..., 10$ & $\pi/3, 2\pi$         & 40    \\
    C &
      \begin{tabular}[c]{@{}l@{}}$-0.1-0.05, ..., 0.2$ \\ and $\pm 0.02$\end{tabular} &
      \begin{tabular}[c]{@{}l@{}}$-10, -5, ..., 20$ \\ and $\pm 2$\end{tabular} &
      $100, 1000$ &
      $3, 8$ &
      $2\pi/3$ &
      324 \\
    D     & $-0.08, -0.02, 0.12, 0.25$ & $-6, 6, 18$                & $80, 111, 167, 400, 2000$ & $4$             & $2\pi/3$              & 60    \\ \hline
    \end{tabular}
\end{table}

The four full factorial parameter sweeps focus on different parameter regimes. 
Sweep~A spans the widest range of shear and veer, while Sweep~B spans the widest range of turbulence intensity and length scales. 
Sweep~C has a high granularity in shear and veer, but over a smaller range of values. 
Finally, Sweep~D has a high granularity and a wide range of Rossby numbers.

\section{Wake comparison between synthetic and ABL profiles}
\label{ssec:results_comparison_TNBL}

In this section, we compare the wake dynamics in the synthetic inflow LES framework introduced in \cref{ssec:methods_les} to simulations of wakes using the concurrent-precursor method \citep{heck_coriolis_2025}, varying the Rossby number between $Ro = 100 \sim 1000$. 
First, in \cref{sssec:tnbl_wake_recovery}, we compare the wake recovery, deflection, and momentum deficit budgets across truly neutral boundary layer (TNBL, also known as the turbulent Ekman layer) inflow conditions and Rossby numbers. 
Then in \cref{sssec:tnbl_L0}, we use the synthetic inflow methodology to examine how integral length scales of turbulence impact wake evolution. 
Finally, in \cref{ssec:results_comparison_SBL}, we compare the synthetic inflow methodology to concurrent-precursor simulations of wakes immersed in stably stratified boundary layer (SBL) inflow conditions.

\subsection{Wake recovery, deflection, and deficit budgets}
\label{sssec:tnbl_wake_recovery}

To verify the proposed synthetic inflow methodology, we simulate a single wind turbine wake in turbulent inflow. 
The concurrent-precursor TNBL simulations in \citet{heck_coriolis_2025} are used as a reference, where the Rossby number $Ro_G = G/(f_c D)$ based on the geostrophic wind speed $G$ varies between $100 \sim 1000$.
In the set of synthetic inflow simulations, labeled ``Synthetic'' throughout this study, the mean wind speed and direction profiles from the TNBL simulations are imposed as inflow conditions in the synthetic inflow method. 
The rotor-averaged $\ti$ is the target in the $\ti$ controller, which matches the inflow turbulence at the rotor location. 
At the rotor plane, the wind veer $\alpha_v$ deviates slightly from the prescribed mean profile due to inflow adjustment.
A summary of the inflow characteristics between the synthetic inflow and TNBL simulations is given in \cref{tab:tnbl_inflows}.

\begin{table}[h]
    \centering
    \caption{Inflow properties of the verification simulations for the TNBL and Synthetic simulations. All simulations are run at a latitude $\phi = \SI{45}{\degree}$.}

    \footnotesize
    \def\arraystretch{1.5}%  1 is the default, change whatever you need
    \begin{tabular}{ll|llllll|llll}
    \hline
    \multicolumn{2}{c|}{\textbf{Shared variables}} &
      \multicolumn{6}{c|}{\textbf{TNBL simulations}} &
      \multicolumn{4}{c}{\textbf{Synthetic inflow simulations}} \\
    Name &
      $Ro$ &
      $Ro_G$ &
      $U_h/G$ &
      $\alpha_v$ &
      $\ti$ (\%) &
      $\psi_h$ ($^\circ$) &
      $z_0$ (m) &
      $\alpha_v$ &
      $\ti$ (\%) &
      $\psi_h$ ($^\circ$) &
      $\lambda_\mathrm{BP}/D$ \\ \hline
    R1000 & 880 & 1000 & 0.87 & 1.4 & 4.3 & -0.5 & $\SI{1e-4}{}$ & 1.3 & 4.4 & -0.5 & $2\pi/3$ \\
    R500  & 461 & 500  & 0.92 & 2.5 & 4.0 & -0.3 & $\SI{1e-4}{}$ & 2.2 & 4.1 & -0.3 & $2\pi/3$ \\
    R250  & 241 & 250  & 0.97 & 4.5 & 3.5 & -0.6 & $\SI{1e-4}{}$ & 4.1 & 3.5 & -0.5 & $2\pi/3$ \\
    R167  & 166 & 167  & 0.99 & 5.6 & 3.2 & -0.4 & $\SI{1e-4}{}$ & 5.3 & 3.3 & -0.4 & $2\pi/3$ \\
    R125  & 125 & 125  & 1.00 & 5.7 & 3.2 & -0.3 & $\SI{1e-4}{}$ & 5.3 & 3.3 & -0.2 & $2\pi/3$ \\
    R100  & 101 & 100  & 1.01 & 5.7 & 3.2 & -0.4 & $\SI{1e-4}{}$ & 5.3 & 3.3 & -0.3 & $2\pi/3$ \\ \hline
    \end{tabular}

    \label{tab:tnbl_inflows}
\end{table}

\begin{figure}[htb]
    \centering
    \includegraphics[width=0.9\linewidth]{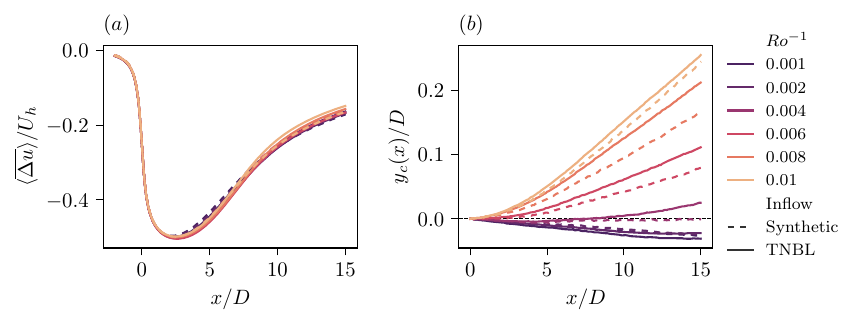}
    \caption{Comparison of ($a$) the streamtube-averaged wake velocity deficit $\langle \overline{\Delta u} \rangle$ and ($b$) wake centroid between the synthetic inflow simulations (dashes) and concurrent-precursor TNBL simulations (solid lines), plotted as a function of the inverse Rossby number $Ro^{-1}$.}
    \label{fig:tnbl_wake_recovery}
\end{figure}

To compare wakes between the concurrent-precursor TNBL inflow and the synthetic inflow, we show the wake recovery and wake deflection in \cref{fig:tnbl_wake_recovery}. 
The wake recovery is computed as the wake velocity deficit $\langle \overline{\Delta u}\rangle$ averaged within the streamtube.
The wake deflection $y_c$ is computed as the geometric centroid of the streamtube position \citep{heck_coriolis_2025}.
Due to wind angle drift in the TNBL, the wind angle at hub height ($\psi_h$), measured in the rotor plane, is not exactly aligned with the rotor. 
Following \citet{heck_coriolis_2025}, the linearized advection $\psi_h \cdot x$ from any misalignment between the inflow angle at the rotor hub height is subtracted off to distinguish inflow-induced wake deflections from mean advection. 
Wake recovery and deflection for the synthetic and TNBL inflow simulations are shown spanning all Rossby numbers in \cref{fig:tnbl_wake_recovery}. 

Overall, wake recovery and deflection agree well between the synthetic inflow and TNBL inflow simulations across all Rossby numbers. 
The normalized mean absolute error (NMAE) between the synthetic and TNBL inflow for the streamtube-averaged wake velocity deficit ranges between $1\%-4\%$. 
Additionally, the magnitude and direction of wake deflections agree between the synthetic and TNBL inflow methods, reaching $\approx 0.2D$ at $Ro = 100$.
In both the synthetic and TNBL inflows, decreasing the Rossby number results in increasingly positive wake deflections monotonically, transitioning from positive to negative wake deflections at $Ro \sim 250$. 
While the quantitative agreement between the TNBL and synthetic inflow-induced wake deflections is lower than the agreement in wake velocity deficit, we note that there is also a considerably larger spread of wake deflections due to unsteadiness in the inflow turbulence over time scales relevant to the ABL and to wind farms \citep{churchfield_using_2016, heck_coriolis_2025}. 

\begin{figure}[htb]
    \centering
    \includegraphics[width=0.95\linewidth]{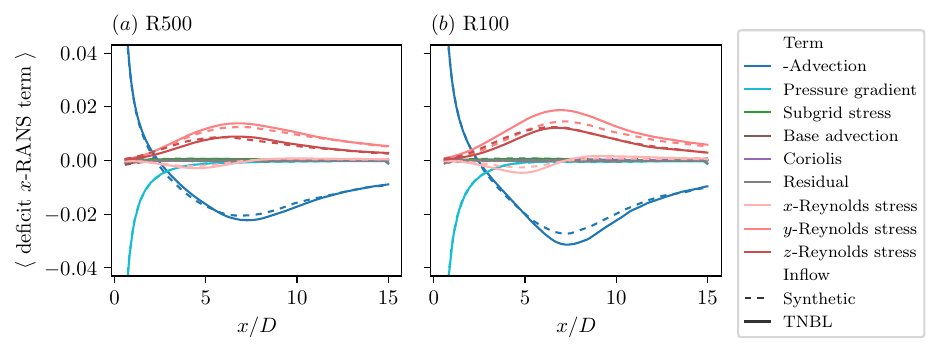}
    \caption{Streamwise evolution of the streamtube-averaged streamwise RANS budget terms for ($a$) relatively weak Coriolis forcing $(\mathrm{R500})$ and ($b$) relatively strong Coriolis forcing $(\mathrm{R100})$. }
    \label{fig:tnbl_streamwise_budget}
\end{figure}

While the wake velocity deficit and deflection agree well between the synthetic and TNBL inflows, it is critical to also compare the individual terms in the momentum balance that govern the wake dynamics. 
For example, error cancellation between terms in the streamwise momentum budget may result in similar wake recovery rates governed by different wake physics. 
In \cref{fig:tnbl_streamwise_budget}, we compare the streamwise momentum deficit terms from \cref{eq:RANS_deficit}, averaged within the streamtube, as a function of streamwise position $x$. 
Budget terms for the $\mathrm{R100}$ and $\mathrm{R500}$ simulations are shown.

The synthetic inflow method matches the streamwise wake dynamics of the concurrent-precursor TNBL simulations.
After the streamwise pressure gradient forcing becomes negligible ($x/D \approx 3$), turbulent momentum entrainment, as noted by the Reynolds stress divergence terms, constitutes $>95\%$ of the wake recovery. 
We further split the Reynolds stress divergence into the three component directions to highlight the importance of vertical and lateral mixing in wake recovery. 
The vertical Reynolds stress gradients match well across all Rossby numbers, which are most strongly influenced by the vertical shear profile in each simulation. 
However, the agreement in lateral Reynolds stress gradients worsens slightly at lower Rossby numbers. 
As we will show in \cref{sssec:tnbl_L0}, the lateral Reynolds stress gradients are strongly influenced by the integral length scale of turbulence. 
The NMAE for the Reynolds stress forcing contributions and advection term is within $10\%$ for all cases, except for the $\mathrm{R100}$ lateral Reynolds stress forcing (shown in \cref{fig:tnbl_streamwise_budget}($b$)), which is underestimated by approximately $20\%$, and which we explore further in \cref{sssec:tnbl_L0}.
Nonetheless, the mean advection and, by extension, wake recovery are well-predicted in all cases (NMAE $< 8\%$). 

\begin{figure}[htb]
    \centering
    \includegraphics[width=0.95\linewidth]{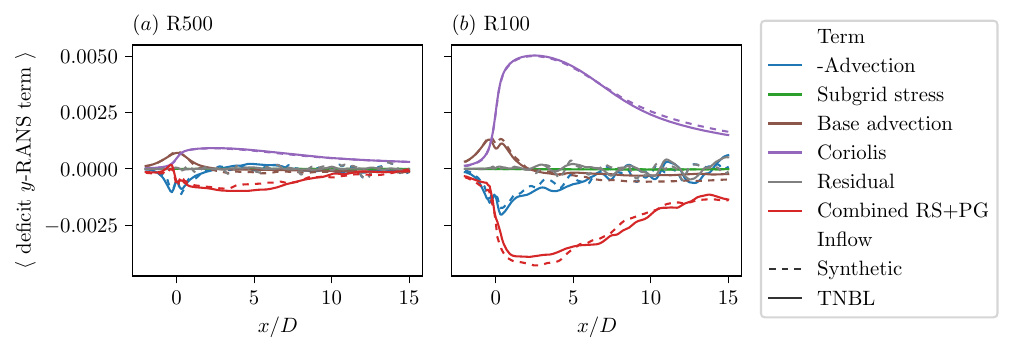}
    \caption{Streamwise evolution of the streamtube-averaged lateral RANS deficit budget terms for ($a$) relatively weak Coriolis forcing $(\mathrm{R500})$ and ($b$) relatively strong Coriolis forcing $(\mathrm{R100})$. }
    \label{fig:tnbl_lateral_budget}
\end{figure}

Finally, we compare the streamtube-averaged forcing terms in the lateral momentum deficit budget, which influence the wake deflection. 
Lateral forcing terms as a function of the streamwise coordinate are shown in \cref{fig:tnbl_lateral_budget}.
For brevity, only the $\mathrm{R500}$ and $\mathrm{R100}$ budgets are shown. 
As with the streamwise momentum deficit budgets, agreement between the synthetic inflow and TNBL wakes is also excellent for the terms in the lateral momentum deficit budget. 
The Coriolis forcing term (purple) in \cref{fig:tnbl_lateral_budget} is matched because the wake velocity deficit is accurately predicted. 
The combined pressure gradient (PG) forcing and Reynolds stress (RS) divergence (red) is also in excellent agreement, and generally opposes the direct Coriolis force. 
Here, we show the sum of the pressure gradient forcing and Reynolds stress terms because a large component of the lateral pressure gradient forcing balances the lateral Reynolds stress divergence (c.f. \citet{pope_turbulent_2000}).
The agreement in wake budgets between the TNBL and the synthetic inflow indicates that the synthetic inflow method can be used to explore wake--ABL interactions and analyze wake dynamics, with comparable accuracy to the more computationally expensive concurrent-precursor method.

\subsection{Turbulence length scale effects}
\label{sssec:tnbl_L0}

In \cref{sssec:tnbl_wake_recovery}, the exact mean wind velocity profiles and hub-height $\ti$ from the TNBL LES were prescribed to the synthetic inflow LES. 
However, the structure of the synthetic turbulence, which is independent of mean velocity and $\ti$ profiles, remains as a free parameter. 
By bandpass-filtering the turbulence in the HIT simulation before it is superimposed into the primary and empty simulations, we can explicitly modify and study the effects of the integral length scale of turbulence on the wake evolution. 
Here, we show two variations of the integral length scale, $L_0^x$ and $L_0^y$, computed with the empty domain hub-height streamwise velocity field $u$ in the streamwise and lateral directions, respectively. 
A summary of the simulations presented in this section is shown in \cref{tab:Ro100_inflows}, alongside streamwise and lateral budget quantities, integrated within the streamtube. 

\begin{table}[htb]
    \centering
    \caption{Simulation parameters of a wind turbine at a turbine diameter-based Rossby number $Ro = 100$, varying the integral length scale. The TNBL case is the same as the $\mathrm{R100}$ case in \cref{tab:tnbl_inflows}.}

    \footnotesize
    \def\arraystretch{1.5}%  1 is the default, change whatever you need

    \begin{tabular}{llllll|ll}
    \hline
    \multicolumn{6}{c|}{Inflow properties}                                                     & \multicolumn{2}{c}{Integrated $x$-RANS budget $\times (D U_h)^{-2}$} \\
    Case label & $\alpha_v$ (deg) & $\ti$ (\%) & $\lambda_\text{BP}/D$ & $L^x_0/D$ & $L^y_0/D$ & $y$-Reynolds stress               & $z$-Reynolds stress              \\ \hline
    \textbf{TNBL} & $5.7$ & $3.3$ & -        & $0.42$ & $0.24$ & 0.059 & 0.044 \\
    Synth-L1      & $5.2$ & $3.4$ & $\pi/2$  & $0.25$ & $0.11$ & 0.045 & 0.041 \\
    Synth-L2      & $5.3$ & $3.3$ & $2\pi/3$ & $0.32$ & $0.18$ & 0.053 & 0.045 \\
    Synth-L3      & $5.3$ & $3.3$ & $\pi$    & $0.46$ & $0.29$ & 0.062 & 0.043 \\
    Synth-L4      & $5.3$ & $3.3$ & $2\pi$   & $0.58$ & $0.41$ & 0.067 & 0.042 \\ \hline
    \end{tabular}

    \label{tab:Ro100_inflows}
\end{table}

\begin{figure}[htb]
    \centering
    \includegraphics[width=0.85\linewidth]{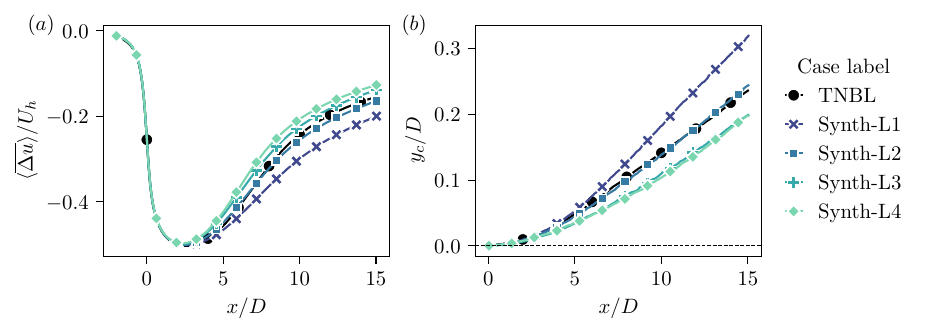}
    \caption{($a$) Wake recovery and ($b$) wake deflection as a function of streamwise position for the inflow profiles approximating the TNBL $\mathrm{R100}$ reference case, varying the turbulent integral length scale $L_0$.}
    \label{fig:R100_recovery_deflection}
\end{figure}

The integral length scale ($L_0$) impacts both wake recovery and wake deflections, shown in \cref{fig:R100_recovery_deflection}. 
The wake recovery rate increases as $L_0$ increases, while the wake deflection magnitude generally decreases as $L_0$ increases.
To understand why the wake recovery is enhanced by increasing $L_0$, we compute the turbulent entrainment of streamwise momentum, integrated within the streamtube from $x\in [0, 15]D$, shown in \cref{tab:Ro100_inflows}.
The vertical Reynolds stress gradient (i.e., turbulent entrainment from aloft) is approximately constant as $L_0$ changes because the inflow wind profile is not affected by $L_0$. 
However, the lateral Reynolds stress gradient (lateral turbulent entrainment of momentum) increases as $L_0$ increases due to enhanced wake meandering \citep{vahidi_influence_2024, bourhis_impact_2025}. 
The closest representation of the wake dynamics in the TNBL is a synthetic inflow simulation that matches $L_0^x$ and $L_0^y$, which would reside between the Synth-L2 and Synth-L3 cases. 
While we can prescribe a high-pass cutoff wavelength of turbulence $\lambda_\mathrm{BP}$, the resulting integral length scales are a function of the flow as well.
For example, the turbulence is elongated in the streamwise direction, resulting in $L_0^x > L_0^y$, despite isotropic bandpass filtering, but the degree of anisotropy varies with $\lambda_\mathrm{BP}$. 
We find that $L_0^y$ is more strongly correlated with $\lambda_\mathrm{BP}$ than $L_0^x$ (see \cref{tab:Ro100_inflows}). 
Future work could improve upon the synthetic turbulence generation and modify the spectral content of the imposed fluctuations. 

Wake deflections decrease with increasing $L_0$ for two reasons. 
First, the direct Coriolis forcing scales linearly with $\overline{\Delta u}$ and contributes to positive (anti-clockwise) wake deflections in the northern hemisphere. 
As a direct result of the enhanced wake recovery, the magnitude of $\overline{\Delta u}$ decreases with increasing $L_0$, and the direct Coriolis force contributes to smaller positive deflections. 
Second, because $v^B$ is negative aloft in positive wind veer, larger length scales of turbulence entrain more negatively flowing lateral momentum from higher aloft. 
As a result, the vertical gradient of $-\Delta \overline{v'w'}$ increases in magnitude as $L_0$ increases, further suppressing positive wake deflections. 
The net deflection is a summation of these competing forcings in the ABL.

\subsection{Wake comparison between synthetic and SBL inflows}
\label{ssec:results_comparison_SBL}

In this section, we provide a brief extension and comparison of the synthetic inflow method to wakes immersed in a stably stratified ABL (SBL). 
The stably stratified ABL simulations are performed using the initialization procedure of \citet{shen_global_2024} with a grid resolution $(\Delta_x, \Delta_y, \Delta_z) = (20, 12.5, 6.25)$~m and $384 \times 256 \times 256$ grid points. 
We perform six simulations of cooling rates between $C_r = 0.1 \sim \SI{0.5}{\kelvin \per \hour}$. 
The SBL simulations have zero free atmosphere stratification following the initialization from \citet{shen_global_2024}. 
All flows are driven by a geostrophic wind speed $G=\SI{12}{\meter \per \second}$ with a roughness length $z_0 = \SI{0.1}{\meter}$. 
The spin-up simulation is performed for $\SI{11}{\hour}$, followed by a short domain rotation to align the hub-height flow with the $x$-axis, and time-averaging is performed from hours $11-20$, following \citet{klemmer_momentum_2024}. 
The mean wind speed, direction, and temperature profiles are then prescribed in the synthetic inflow simulation, and a Dirichlet boundary condition is used to enforce a constant ground temperature. 
A summary of inflow characteristics for the SBL simulations and the stratified synthetic inflow simulations is given in \cref{tab:sbl_inflows}

\begin{table}[htb]
    \caption{Parameters for SBL and stratified synthetic inflow simulations. All simulations are run at a latitude $\phi=45^\circ$, and the turbine is placed at $z_h/D=0.625$.}
    \centering
    \footnotesize
    \def\arraystretch{1.5}%  1 is the default, change whatever you need
    
    {\setlength{\tabcolsep}{4pt}
    \begin{tabular}{lllllll|lllllll}
    \hline
    \multicolumn{7}{c|}{\textbf{SBL simulations}}         & \multicolumn{7}{c}{\textbf{Synthetic inflow simulations}} \\
    $C_r$ ($\SI{}{\kelvin\per\hour})$ &
      $U_h/G$ &
      $\ti$ (\%) &
      $\alpha_v$ ($^\circ$) &
      $\psi_h$ ($^\circ$) &
      $L_0^x/D$ &
      $L_0^y/D$ &
      $U_h/G$ &
      $\ti$ (\%) &
      $\alpha_v$ ($^\circ$) &
      $\psi_h$ ($^\circ$) &
      $L_0^x/D$ &
      $L_0^y/D$ &
      $\lambda_\mathrm{BP}/D$ \\ \hline
    0.1 & 0.81 & 5.2 & 11.8 & -0.9 & 0.26 & 0.10 & 0.80 & 5.2 & 11.1 & -1.4 & 0.29 & 0.11 & $\pi/4$ \\
    0.2 & 0.83 & 4.6 & 14.8 & -1.2 & 0.22 & 0.08 & 0.83 & 4.6 & 14.2 & -1.8 & 0.21 & 0.10 & $\pi/6$ \\
    0.3 & 0.85 & 4.2 & 18.3 & -1.5 & 0.18 & 0.07 & 0.85 & 4.1 & 17.7 & -2.2 & 0.22 & 0.08 & $\pi/8$ \\
    0.4 & 0.87 & 3.8 & 23.4 & -2.1 & 0.17 & 0.07 & 0.88 & 3.8 & 22.9 & -2.9 & 0.17 & 0.07 & $\pi/8$ \\
    0.5 & 0.90 & 3.4 & 28.2 & -2.4 & 0.15 & 0.06 & 0.91 & 3.4 & 27.6 & -3.1 & 0.18 & 0.07 & $\pi/8$ \\ \hline
    \end{tabular}
    }
    \label{tab:sbl_inflows}
\end{table}

\begin{figure}[htb]
    \centering
    \includegraphics[width=0.9\linewidth]{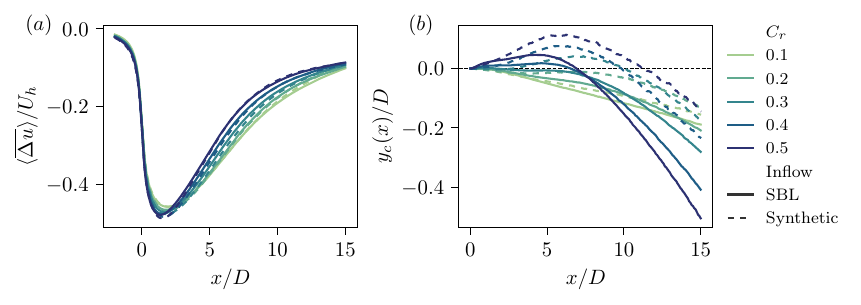}
    \caption{($a$) Wake recovery and ($b$) wake deflection as a function of streamwise distance compared between SBL and stratified synthetic inflow LES for varying cooling rates $C_r$ (in $\SI{}{\kelvin \per \hour}$, see \cref{tab:sbl_inflows}).}
    \label{fig:sbl_wakes}
\end{figure}

In \cref{fig:sbl_wakes}($a$), we compare wake recovery in the SBL simulations with the wake recovery from the stratified synthetic inflow simulations, quantified with the streamtube-averaged velocity deficit. 
Overall, the wake recovery agrees between the methods, even at large cooling rates. 
Interestingly, despite the decreasing length scales of turbulence $L_0$ and decreasing $\ti$ at higher cooling rates, wake recovery tends to increase with increasing $C_r$. 
We attribute this to enhanced wake recovery resulting from increasing wind veer across the rotor ($\alpha_v$), which is discussed in-depth in \cref{ssec:explore_recovery}. 

In contrast with the relatively strong agreement in wake recovery between the stratified synthetic simulations and the SBL reference cases, the quantitative predictions of wake deflections are less strong. 
The qualitative shape and trends are captured, showing an initially positive wake deflection in the near-wake, which becomes negative in the turbulent far-wake. 
The quantitative mismatch in wake deflection between the stratified synthetic inflow simulations and the SBL simulations is driven by several factors. 
First, the magnitude of terms in the lateral momentum budget is much smaller than in the streamwise momentum budget, taking longer to converge. 
We opt to time-average from hours $11-20$ in all cases, rather than a full inertial period, due to the transience in the stable boundary layer.
As a result, the reference SBL wake deflection contains inherent variability from inertial oscillations. 
Second, the SBL simulations evolve with time, while the synthetic inflow simulations are statistically quasi-steady. 
Here, we are averaging over changes in wind direction in the SBL, which drifts due to surface cooling \citep[c.f.,][]{shen_global_2024}. 
By contrast, the synthetic inflow case is statistically stationary, which is in part an advantage as it enables computing well-converged turbulence statistics. 

In summary, the synthetic inflow method provides a simplified, controllable framework for investigating stratified ABL effects on wake dynamics, reproducing the flow physics of concurrent-precursor LES. 
Furthermore, we can use the methodology to study the effects of turbulence integral length scales, which do not affect profiles of wind speed, direction, or $\ti$, on wake dynamics. 
These results highlight an opportunity for studying stratified wake dynamics in a controlled environment unaffected by SBL transience, which we suggest exploring in future work. 
Replacing the ABL mean wind profiles from concurrent-precursor simulations with the simplified inflow parameterization given in \cref{eq:U_phi} also captures correct parametric trends in wake recovery and deflection, shown in \cref{appx:linear_profiles}, which we will use to explore a wide range of ABL parameters.

\section{ABL parameter exploration}
\label{sec:results_exploration}
In this section, we use the methodology verified in \cref{ssec:results_comparison_TNBL} to sweep over a wide range of wind shear, wind veer, $\ti$, Rossby numbers, and turbulence length scales to explore the effects of ABL parameters on wake development. 
In total, we analyze 640 new LES cases that encompass the parameter space of ABL flow conditions observed in field measurements using the parameter sweeps described in \cref{ssec:methods_mvco}.
Beginning in \cref{ssec:explore_recovery}, we investigate the effects of the ABL inflow on wake recovery using the streamwise momentum budget. 
Then, in \cref{ssec:explore_deflection}, we examine ABL-induced wake deflections and identify regimes of wake deflections resulting from Coriolis forces and ABL interactions. 

\subsection{Wake recovery regimes and streamwise dynamics}
\label{ssec:explore_recovery}

\begin{figure}[htb]
    \centering
    \includegraphics[width=1\linewidth]{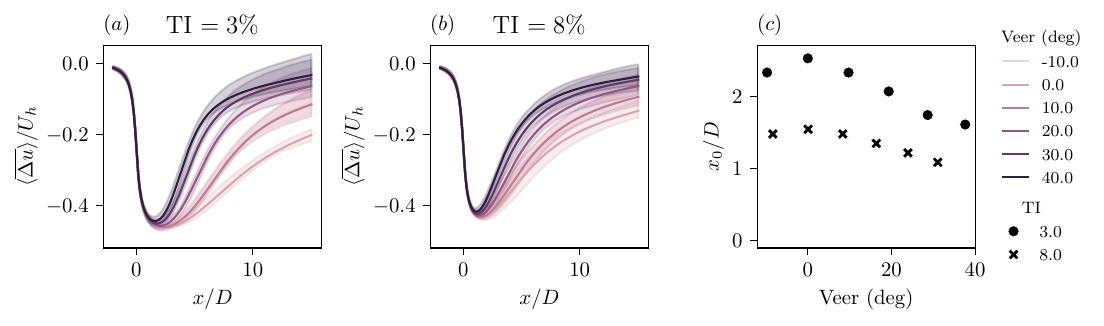}
    \caption{Streamtube-averaged wake velocity deficit as a function of downstream distance for ($a$) $\ti = 3\%$ and ($b$) $\ti = 8\%$. In ($a$, $b$), the shaded region shows the full range of wake recovery across different values of shear and Rossby number. ($c$) Near-wake length $x_0$ computed with the ensemble mean velocity deficit as a function of veer and $\ti$. }
    \label{fig:ABL_recovery}
\end{figure}

To begin, we investigate the effect of the ABL inflow on wind turbine wake recovery. 
We begin with data from Sweep~A in \cref{tab:LES_inputs}, covering the widest range of wind shear and wind veer. 
The streamtube-averaged wake velocity deficit is shown in \cref{fig:ABL_recovery} as a function of downstream distance. 
Within the ranges of parameters observed in the field, veer and inflow turbulence intensity have the largest effect on the wake recovery rate. 
The influence of wind veer on wake recovery is largest at low $\ti$.
At high $\ti$, wake recovery is generally faster for all cases than at lower $\ti$, and the effects of wind veer are still present but diminished. 
The shaded regions in \cref{fig:ABL_recovery}($a$-$b$) show the spread in wake recovery due to the full variation in wind shear and $Ro$.

As the magnitude of the inflow wind veer increases, two effects synergistically enhance wake recovery. 
First, the near-wake length $x_0$ (i.e., the $x$-location of minimum wake velocity) decreases with increasing wind veer magnitude, as shown in \cref{fig:ABL_recovery}($c$).
The dependence of $x_0$ on wind veer is stronger at low inflow $\ti$, and is symmetrical across positive wind veer and negative wind veer (also called backing, where winds turn to the left with increasing height in the northern hemisphere). 
As a result, the wake recovery begins closer to the rotor disk, and at a given point downstream, wake recovery is enhanced because the far wake has been turbulent for a longer distance. 
Second, the rate (slope) of wake recovery increases with increasing wind veer magnitude, which has also been observed in concurrent-precursor simulations \citep{abkar_influence_2016, churchfield_effects_2018}. 

\begin{figure}[htb]
    \centering
    \includegraphics[width=1\linewidth]{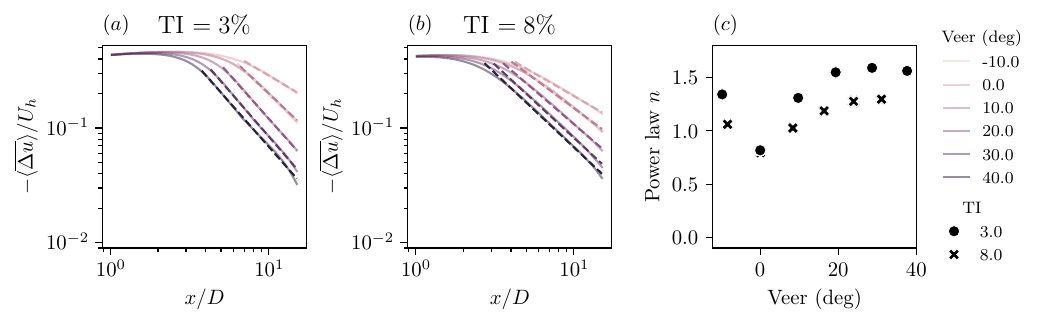}
    \caption{($a$, $b$) Negative of \cref{fig:ABL_recovery}($a$, $b$) except in log-log axes. Dashed lines are power law fits to the ensemble mean (across shear and Rossby number) velocity deficit profiles as a function of veer and $\ti$. ($c$) power law decay exponent computed by power law fits to ensemble mean profiles.}
    \label{fig:ABL_recovery_loglog}
\end{figure}

Transforming the wake recovery profiles into log-log space elucidates the accelerated rate of wake recovery due to wind veer. 
This is shown in \cref{fig:ABL_recovery_loglog}($a$, $b$). 
Empirically, we can approximate wake recovery rate using a power law fit $\langle \overline{\Delta u} \rangle / U_h = a (x/D)^{-n}$, where $n$ is the power law decay exponent and $a$ is a scaling factor \citep{vermeulen_experimental_1980, hogstrom_field_1988, stevens_flow_2017}. 
Here, the exponent $n$ is fit to the velocity deficit profile beyond the point where the concavity of the velocity deficit profile changes, which has been described as the start of the fully turbulent wake \citep{fei_analytical_2025}. 
We fit to the ensemble mean across all shears and Rossby numbers (solid lines in \cref{fig:ABL_recovery}($a$, $b$)) for simplicity. 
The resulting power law decay exponents are shown in \cref{fig:ABL_recovery_loglog}($c$), where increasing the power law wake decay exponent increases symmetrically with increasing veer magnitude.
The range of power law wake decay rates that we fit from the synthetic inflow simulations falls within the range of reported values from wind tunnel experiments \citep{chu_turbulence_2014} and in field campaigns \citep{hogstrom_field_1988, iungo_volumetric_2014}. 

\begin{figure}[htb]
    \centering
    \includegraphics[width=1\linewidth]{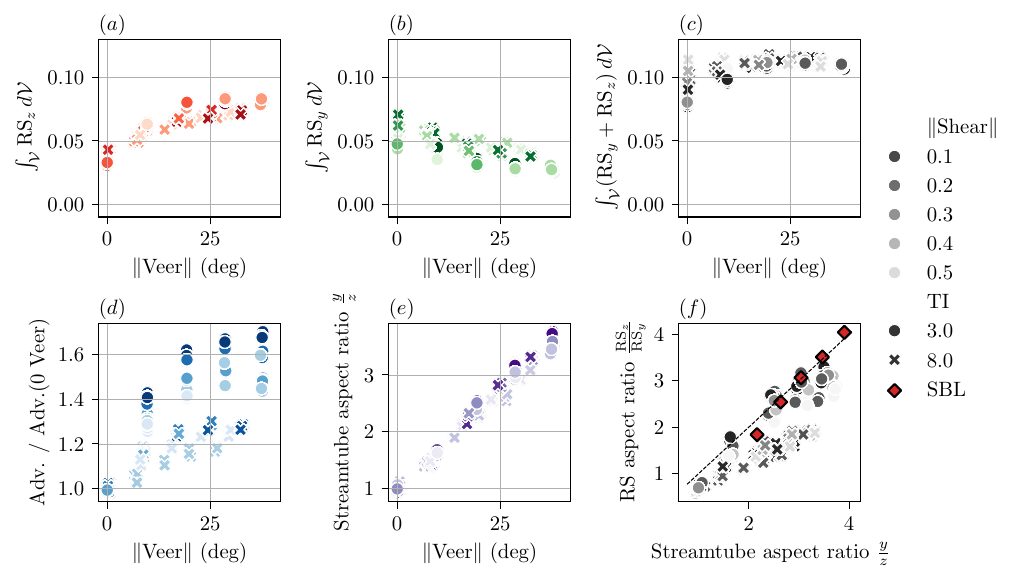}
    \caption{($a$--$c$): Streamtube-integrated Reynolds stress contributions as a function of wind veer at the rotor from the ($a$) vertical and ($b$) lateral directions, and ($c$) combined. ($d$) Streamtube-integrated advection, relative to the equivalent inflow but without veer, versus mean streamtube aspect ratio. ($e$) Correlation of wind veer and mean streamtube aspect ratio. ($f$) Correlation of the ratio of vertical/lateral Reynolds stress contributions to the mean aspect ratio of the streamtube $y/z$.}
    \label{fig:du_dynamics}
\end{figure}

To explain the increasing wake recovery with increasing veer magnitude, we analyze the streamwise momentum deficit budget. 
Specifically, the wake velocity deficit is the integral of the advection term \citep{heck_coriolis_2025}. 
In \cref{fig:du_dynamics}, we show the parametric dependence of the integrated Reynolds stress contributions within the streamtube from the rotor to $x/D=15$ downstream. 
As the wind veer magnitude increases, the total vertical entrainment of momentum increases (\cref{fig:du_dynamics}($a$)), while the lateral entrainment of momentum only slightly decreases (\cref{fig:du_dynamics}($b$)). 
The sum of the Reynolds stress divergence terms, shown in \cref{fig:du_dynamics}($c$), increases with increased inflow wind veer magnitude. 
Relative to the zero veer case, the increased turbulent flux of momentum increases the mean advection with increasing veer magnitude, shown in \cref{fig:du_dynamics}($d$).
To summarize, the enhanced rate of wake recovery in veered inflow conditions, quantified by the power law decay exponent $n$, is caused by an increase in turbulent momentum entrainment from aloft. 

It is well established that wakes in veered conditions will elongate and skew as different vertical levels of the wake are exposed to changing directions of inflow \citep{magnusson_influence_1994, abkar_influence_2016}. 
The height of the wake remains nearly the same, but the wake is elongated in the lateral direction. 
Here, we measure the aspect ratio of the wake as the ratio of the maximum extents of the streamtube in the lateral to the vertical directions. 
Due to wake skewing, which is strongly correlated with the inflow veer (\cref{fig:du_dynamics}($e$)), the elongated wakes have a greater surface area. 
This allows for increased turbulent entrainment of momentum from above. 
In \cref{fig:du_dynamics}($f$), we compare the ratio of integrated vertical to lateral momentum entrainment (labeled $\mathrm{RS}_z/\mathrm{RS}_y$) to the streamtube aspect ratio $y/z$. 
For the $\ti=3\%$ simulations and the concurrent-precursor SBL simulations in \cref{ssec:results_comparison_SBL}, the correlation between the geometric aspect ratio of the wake and the ratio of turbulent stresses is nearly 1:1, indicating that the geometric effect of wake skewing enhances wake recovery. 

\begin{figure}[htb]
    \centering
    \includegraphics[width=0.9\linewidth]{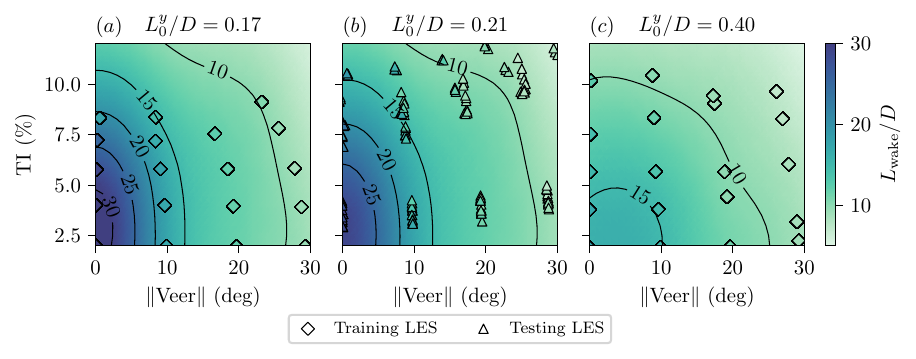}
    \caption{Regimes of wake length $L_\mathrm{wake}$ (distance where the wake recovers to $25\%$ of the maximum deficit magnitude) plotted as a function of veer and hub height turbulence intensity, parsed between ($a$) relatively small integral length scale inflow ($L^y_0/D = 0.17 \pm 0.03$), ($b$) out-of-sample testing ($R^2 = 0.86$), and ($c$) large integral length scale inflow ($L^y_0/D = 0.40 \pm 0.05$). Contours are interpolated using Gaussian process regression.}
    \label{fig:du_contours}
\end{figure}

In this exploration, we have observed that wind turbine wake recovery is most strongly influenced by the inflow wind veer and the turbulence intensity. 
Additionally, from \cref{sssec:tnbl_L0}, the integral length scale of turbulence also has a leading-order role in wake recovery. 
To conclude this section, we explore parameter regimes of wake recovery using the synthetic inflow methodology to sweep over ranges of the three variables with the largest influence on wake recovery: wind veer ($\alpha_v = 0 \sim 30^\circ$, $\ti = 2\% \sim 10\%$, and integral length scale ($\lambda_\text{BP} = \pi/3, 2\pi$).
The exact simulation inputs are given in Sweep~B in \cref{tab:LES_inputs}.
Because the inflow properties influence the wake dynamics in several ways (near-wake length, power law decay constant, etc.), we introduce a simple metric that encodes all of these effects into a practical measure of wake recovery: the distance $L_\text{wake}$ at which the wake velocity deficit $\langle \overline{\Delta u}\rangle$ recovers to $25\%$ of its maximum value in magnitude. 
For cases where the wake velocity deficit magnitude does not reach $25\%$ of its maximum value inside the simulation domain, a power law fit is used to extrapolate the value of $L_\text{wake}$, which is justified by the goodness-of-fit in \cref{fig:ABL_recovery_loglog}($a$, $b$). 
Contours of wake recovery regimes are shown in \cref{fig:du_contours}, where a Gaussian process (GP) regressor is used to interpolate between the LES data points using $\alpha_v$, $\ti$, and $L_0^y$ as inputs.
The LES cases from Sweep~A are used in the out-of-sample evaluation for the GP accuracy ($R^2 = 0.86$). 

The comprehensive simulations performed here allow the wake recovery to be mapped across the atmospheric conditions in which wind turbines will operate. 
In \cref{fig:du_contours}, we study three key ABL flow variables for wake recovery: 1) $\ti$; 2) integral length scale of turbulence; and 3) wind veer.
At lower $L_0/D$, the wake length is more sensitive to wind veer and $\ti$ than at high $L_0/D$. 
Similarly, at low $\ti$, wind veer has a larger impact on the wake length than at high $\ti$.
We can contextualize these trends using specific turbine models in varying inflow conditions. 
For example, a large offshore wind turbine like the IEA~15~MW reference turbine ($D \sim \SI{250}{\meter}$) may experience a relatively small $L_0/D$ due to its size and operate in relatively low inflow turbulence intensities \citep[c.f.,][]{bodini_offshore_2020}). 
From \cref{fig:du_contours}($a$), the wake of these conditions is significantly influenced by the magnitude of wind veer in the inflow: varying the veer magnitude from $0^\circ$ to $12^\circ$ halves the wake length. 
By contrast, a smaller \SI{100}{\meter} wind turbine will operate at relatively larger $L_0/D$, where the effects of wind veer are significantly diminished, regardless of $\ti$, shown in \cref{fig:du_contours}($c$). 
The interplay between the turbulence length scale, $\ti$, and wind veer depends not only on the turbine size but also on the instantaneous inflow conditions. 
Quantifying the impact on wake recovery within the complex, multi-dimensional ABL parameter space elucidates key relationships between the wake evolution and the properties of the incident inflow, which may inform future wind farm design and control.

\subsection{Wake deflection regimes and lateral dynamics}
\label{ssec:explore_deflection}

\begin{figure}[htb]
    \centering
    \includegraphics[width=1\linewidth]{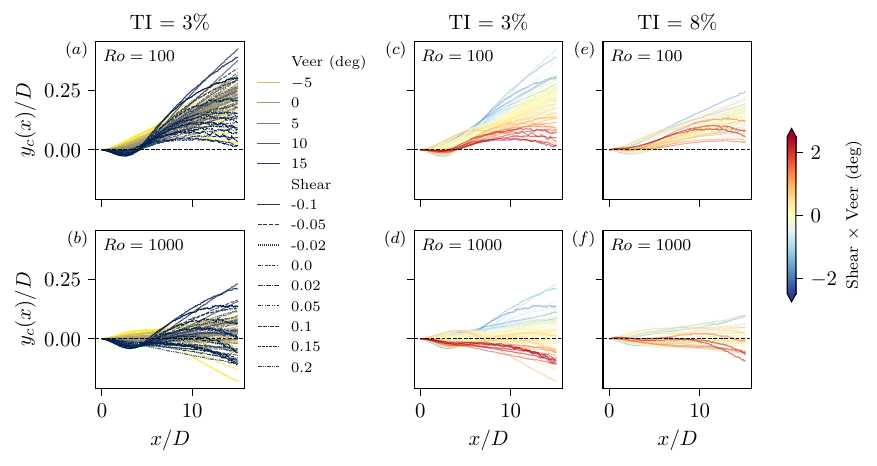}
    \caption{Wake deflections for a fine sweep of shear ($\alpha_s = -0.1 \sim 0.2$) and veer ($\alpha_v = \SI{-10}{\degree} \sim \SI{20}{\degree}$) at $3\%$ and $8\%$ turbulence intensity. ($a$, $c$, $e$) Deflections at relatively strong Coriolis forcing, $Ro = 100$, ($b$, $d$, $f$) deflections at relatively weak Coriolis forcing, $Ro = 1000$. ($a$, $b$) Wake deflections colored by the wind veer $\alpha_v$. ($c$--$f$) Wake deflections colored by the product of shear and veer, $\sxv$.}
    \label{fig:ABL_deflections}
\end{figure}

In this section, we investigate wake deflections induced by Coriolis effects, which depend parametrically on incident conditions. 
We simulate wakes for Sweep~C inflows (\cref{tab:LES_inputs}) over shear ($\alpha_s = -0.1 \sim 0.2$), veer ($\alpha_v = -10^\circ \sim 20^\circ$), $\ti =3\%, 8\%$, and Rossby number ($Ro = 100, 1000$), at higher granularity in shear and veer than Sweep~A used in \cref{ssec:explore_recovery}. 
Wake deflections ($y_c$) as a function of streamwise distance are shown in \cref{fig:ABL_deflections}.

Wake deflections induced by the simulated inflow conditions may be positive or negative in sign. 
In \cref{fig:ABL_deflections}($a$, $b$), wake deflection trajectories are split by Rossby number and colored by the veer strength $\alpha_v$, while the linestyle is varied with the shear strength $\alpha_s$. 
There is a net shift toward increasingly positive wake deflections across all shear and veer conditions from the high Rossby number (weak Coriolis) inflows to the low Rossby number (strong Coriolis) inflows. 
Notably, a wide spread of wake deflections may be observed for a fixed amount of inflow veer, due to varying inflow shear, as noted by the variation in $y_c$ for a given hue. 
The same is also true for shear, where varying the inflow veer also leads to a spread of values $y_c$ for a fixed amount of wind shear. 

Rather than parsing the effects of shear and veer separately, figures \cref{fig:ABL_deflections}($c$--$f$) show wake deflections colored with the product of inflow shear and veer ($\sxv$), which elucidates a novel scaling relationship between the inflow and the observed deflection. 
The observed wake deflections are strongly a function of this combined parameter rather than shear and veer independently, particularly at $\ti = 3\%$. 
At $\ti=8\%$, we observe that the collapse of wake deflections worsens, indicating that knowledge of shear and veer separately is important to the lateral wake dynamics as $\ti$ increases. 

\begin{figure}[htb]
    \centering
    \includegraphics[width=1\linewidth]{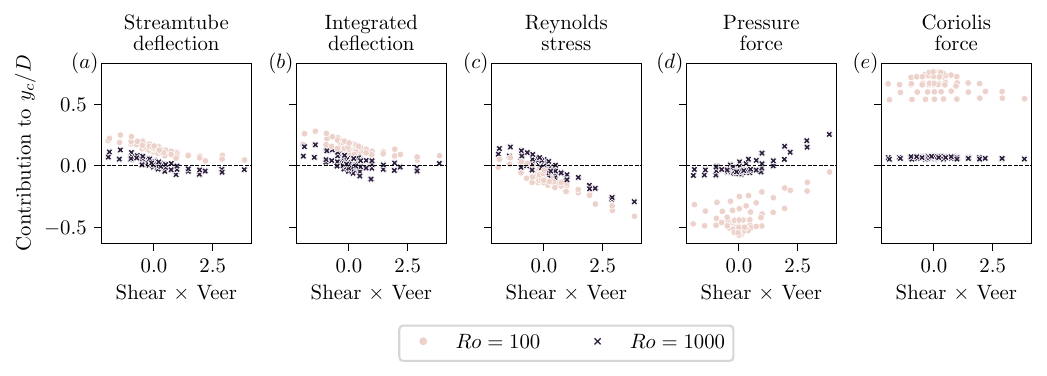}
    \caption{Lateral wake deflection and dynamics, plotted against the product of the local inflow shear and veer at the rotor $\sxv$ for an inflow $\ti=3\%$ and $x/D=10$. ($a$) Wake deflection computed as the centroid of the streamtube seeded at the rotor. ($b$) The integrated advection and base advection terms equal the observed wake deflection. ($c$--$e$) Streamtube-averaged budget terms integrated using \cref{eq:rans_integrated} showing individual forcing contributions to the observed deflection from ($c$) lateral Reynolds stresses, ($d$) lateral pressure forcing, and ($e$) direct Coriolis forcing.}
    \label{fig:deflection_budgets}
\end{figure}

To investigate the cause of the collapse in the wake deflections, we consider the lateral deficit RANS equations (\cref{eq:RANS_deficit}). 
As is done in \citet{heck_coriolis_2025}, we can reconstruct the individual forcing contributions to the net wake deflection by integrating the forcing terms in the lateral momentum budget ($f_{y,j}$) in a Lagrangian frame of reference: 
\vspace{-2pt}
\begin{equation}
\label{eq:rans_integrated}
    y_c(x) = \sum_j \int_0^x
    \frac{1}{\langle \overline{u} \rangle(x')}
    \int_0^{x'}
    \frac{1}{\langle \overline{u} \rangle(x'')} \langle \overline{f}_{y, j} \rangle(x'')
    \,dx''
    \,dx'. 
\end{equation}
The observed wake deflection is the integral of the advection term in the full lateral RANS budget $\bar{u}_j \cdot \tfrac{\partial \overline{v}}{\partial x_j}$.
Assuming the advection term in the background flow to be negligible ($\overline{u_j^B} \cdot \tfrac{\partial \overline{v^B}}{\partial x_j} = 0$, e.g., horizontally homogeneous flow), the full RANS advection term is equal to the sum of the advection and base advection terms in the deficit RANS budget in \cref{eq:RANS_deficit}. 
In \cref{fig:deflection_budgets}($a$), the measured wake deflection at $x/D = 10$ downstream is shown as a function of $\sxv$ and the Rossby number at $\ti = 3\%$, which matches the sum of the advection and base advection terms (labeled ``Integrated deflection'' in \cref{fig:deflection_budgets}($b$)).
We show the contributions from the Reynolds stresses, lateral pressure forcing, and Coriolis forcing in \cref{fig:deflection_budgets}($c$--$e$), integrated within the streamtube from the rotor to $x/D = 10$. 

The explanation for the collapse of the wake deflection to the axis $\sxv$ is given by the individual momentum budget terms. 
Specifically, the wake deflection contribution from the Reynolds stress divergence, shown in \cref{fig:deflection_budgets}($c$), collapses with an approximately linear scaling when plotted against the product $\sxv$.
The lateral pressure force balances the dominant forcing in the flow: at low $\sxv$, the pressure opposes the Coriolis forcing, but as $\sxv$ increases, the pressure also collapses with approximately linear scaling as a function of $\sxv$. 
That is, the sum of the lateral pressure force and Coriolis force scales with $\sxv$ (not shown), even though the Coriolis contribution to $y_c$ instead scales with veer alone. 
Overall, these lateral dynamics contribute to the collapse of wake deflections onto the combined variable $\sxv$, the product of shear and veer, particularly at low $\ti$. 

\begin{figure}[htb]
    \centering
    \includegraphics[width=1\linewidth]{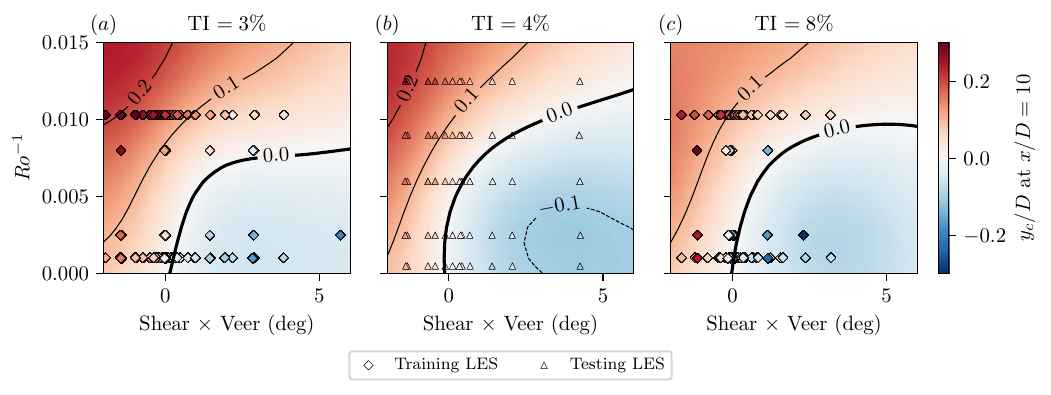}
    \caption{Contours of wake deflection at $x/D=10$ as predicted by the Gaussian process regressor with LES points overlaid, for ($a$) $\ti = 3\%$, ($b$) $\ti = 4\%$ (out-of-sample), and ($c$) $\ti = 8\%$. Coriolis forcing strength increases as $Ro^{-1}$ increases. }
    \label{fig:deflection_contours}
\end{figure}

Finally, we identify regimes of wake deflections as a function of inflow parameters using the synthetic turbulent inflow methodology. 
Using the wide range of ABL parameters in Sweep~A and Sweep~C, totaling 540 simulations, we train a GP to estimate the wake deflection as a function of the combined shear parameter $\sxv$, $Ro$, and $\ti$. 
Wake deflection contours are shown at varying turbulence intensities in \cref{fig:deflection_contours}. 
A separate testing dataset (Sweep~D in \cref{tab:LES_inputs}) is used to evaluate the efficacy of the GP ($R^2 = 0.82$), shown in \cref{fig:deflection_contours}($b$). 

The contours delineate an interface between positive and negative wake deflections. 
Positive $\sxv$ corresponds with the canonical Ekman spiral, thus ABL winds tend to induce negative wake deflections in the northern hemisphere. 
As the Rossby number decreases, direct Coriolis forces become increasingly relevant, and wake deflections become increasingly positive.
The critical Rossby number, which delineates the boundary between positive and negative deflections, depends on the incident flow. 
In general, higher wind shear and wind veer decrease the critical Rossby number. 
Note, however, that negative combinations of $\sxv$ also contribute to positive wake deflections. 
That is, if either the wind shear $\alpha_s$ is negative, or the wind veer $\alpha_v$ is negative---but not both---then the shear and veer-driven deflection can work with the direct Coriolis forcing, resulting in large positive deflections.
Finally, at zero wind shear \textit{or} zero wind veer, and for $Ro^{-1} \to 0$ (i.e., negligible Coriolis forcing), no wake deflections are expected. 
That is, wind veer in the absence of wind speed shear (or vice-versa) is insufficient to create inflow-induced wake deflections. 
At elevated $\ti$, the collapse of wake deflections with the combined shear parameter $\sxv$ is weaker, as highlighted by \cref{fig:ABL_deflections}($e$, $f$). 
However, for brevity, we do not include the separated effects of wind shear and wind veer on wake deflections here. 

The wake deflection regimes identified by the synthetic inflow methodology may lead to an improved understanding of wake deflection mechanisms and modeling. 
Coriolis-induced lateral wake deflections can cause an asymmetry in the power production of downwind turbines \citep{bromm_numerical_2017} and also affect the optimal wind farm control strategy \citep{nouri_coriolis_2020}. 
However, modeling inflow-induced deflections is challenging due to the influence of competing forces, including pressure gradients and turbulence entrainment. 
Simplified scaling relationships, such as the collapse of wake deflections onto the product of shear and veer $\sxv$, are useful for guiding improvements in wake modeling that can inform decisions on wind farm design and control.

\section{Conclusion and future work}
\label{sec:conclusions}

We propose a synthetic inflow large eddy simulation method for studying the influence of ABL inflow properties on wind turbine wakes. 
By using slip walls, synthetic turbulence generation, prescribed inflow profiles, and a modified geostrophic wind forcing, the synthetic inflow method allows for independent controllability of wind shear, veer, inflow turbulence intensity, and Coriolis forcing (denoted by the Rossby number $Ro = U_h/(f_c D)$).
In contrast to the concurrent-precursor LES method, where inflow variables are coupled through ABL physics, the synthetic inflow method disentangles the inflow variables, enabling targeted experiments on wake physics. 
We verify that the synthetic inflow method through comparisons to truly neutral and stably stratified ABL simulations. 
The synthetic inflow method recreates the correct wake recovery within 4\% across all cases and captures the correct streamwise wake dynamics. 
Qualitatively, trends in wake deflection agree between LES methods, but discrepancies are larger in the wake deflection than in the wake recovery due to the sensitivity to inflow direction arising from transience in the ABL, which is not present in the synthetic inflow method.
Unlike concurrent-precursor ABL simulations, the synthetic inflow method is statistically stationary, which is advantageous for computing well-converged turbulence statistics. 

Using the proposed synthetic inflow method, we investigate the effects of inflow winds on wake dynamics across a wide range of shear ($\alpha_s$), veer ($\alpha_v$), Coriolis forcing magnitudes, and turbulence intensities and length scales. 
Parameter ranges were selected based on lidar field observations on the north Atlantic coast of the United States. 
We find that wind veer, inflow turbulence intensity, and turbulence length scales have the largest impact on wind turbine wake recovery. 
Specifically, increasing $\ti$ and increasing turbulence length scales enhance wake recovery through increased turbulent mixing and meandering, respectively, echoing conclusions from previous studies \citep{wu_atmospheric_2012, du_influence_2021}.
Increasing wind veer magnitude also enhances wake recovery. 
We identify that the important role of veer on wake recovery is driven by an increase in wake surface area and, consequently, an increased turbulent momentum flux into the wake. 
We find that for wake deflections, Rossby number, turbulence intensity, and the product of wind shear and wind veer $\sxv$ are the key parameters governing lateral wake dynamics and corresponding wake deflection from Coriolis effects. 
In particular, the turbulent entrainment of lateral momentum collapses when plotted against the product $\sxv$, indicating that both wind shear \textit{and} wind veer are needed to create the asymmetry for turbulence-driven lateral wake deflections; wind veer in isolation is not sufficient to cause wake deflections.

Several unexplored parameter axes of ABL inflow can easily be integrated into the synthetic inflow framework. 
A deliberate exploration of stratified turbulence, including anisotropic synthetic inflow turbulence, would be useful in clarifying the role of thermal stratification on wake dynamics. 
Furthermore, the spectral structure of the inflow turbulence may affect wake evolution, as suggested by other studies \citep[e.g.,][]{tobin_modulation_2019, gambuzza_influence_2023}.
Alternative methods of synthetic turbulence generation \citep[e.g.,][]{mann_wind_1998, poletto_new_2013} may further simplify the method and reduce computational cost \citep{meneveau_big_2019}. 

The findings from the synthetic inflow method can be used to complement concurrent-precursor LES with all of the coupled ABL physics included. 
For example, in stable boundary layers, there is a trade-off between increasing wind veer, which enhances wake recovery, but decreasing $\ti$ and length scales, which diminishes wake recovery. 
Separately, larger wind turbines will operate higher in the ABL \citep{wu_near-ground_2021}, decreasing $\ti$ and therefore wake recovery rates, but this effect may be offset by larger inflow length scales and higher magnitudes of shear and veer. 
As contemporary-scale wind turbines are constructed and installed in new ABL regimes, improving our understanding of flow physics across ABL conditions may improve wake modeling and lead to increases in the efficiency of wind energy installations.

\paragraph{Acknowledgements} 
Simulations were performed on the Stampede3 and Anvil supercomputers under the NSF ACCESS project ATM170028.

\paragraph{Funding Statement}
K.S.H. and M.F.H. gratefully acknowledge funding from the National Science Foundation (Fluid Dynamics program, grant number FD-2226053, Program Manager: Dr. Ronald D. Joslin).
K.S.H. acknowledges additional funding through a National Science Foundation Graduate Research Fellowship under grant no. DGE-2141064. 

\paragraph{Declaration of Interests}
The authors declare no conflicts of interest.

\paragraph{Author Contributions}
M.F.H. and K.S.H. conceived the research. 
K.S.H. developed the code for the methodology, performed the large eddy simulations, and analyzed the data.
All authors contributed to manuscript writing and edits.

\paragraph{Data Availability Statement}
Data that support the findings of this study are available from the authors upon request.

% \paragraph{Ethical Standards}
% The research meets all ethical guidelines, including
% adherence to the legal requirements of the study country.

% \paragraph{Supplementary Material}

\begin{appendix}
\section{Parameterized shear and veer approximation}
\label{appx:linear_profiles}
\begin{figure}[htb]
    \centering
    \includegraphics[width=0.9\linewidth]{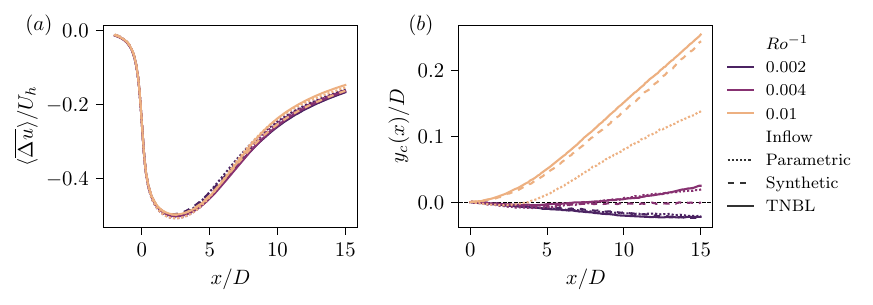}
    \caption{Comparison of ($a$) the streamtube-averaged wake velocity deficit and ($b$) wake centroid between concurrent-precursor TNBL simulations (solid), synthetic inflow simulations with TNBL inflow profiles (dashed), and synthetic inflow simulations with best-fit parameterized profiles (dotted) given by \cref{eq:U_phi}.}
    \label{fig:wakes_linear_tnbl}
\end{figure}

In this appendix, we compare the wake evolution using the parameterized representation of the ABL inflow velocity profiles to the wake evolution using the prescribed mean profiles from the truly neutral ABL simulations in the synthetic inflow method. 
The wind shear $\alpha_s$ and wind veer $\alpha_v$ parameters are fit to the ABL velocity profiles given by \cref{eq:U_phi} within the rotor extent.
The resulting wake recovery and wake deflection are shown in \cref{fig:wakes_linear_tnbl}, comparing the concurrent-precursor TNBL inflow, synthetic inflow with mean TNBL inflow profiles (``Synthetic''), and synthetic inflow with parameterized inflow profiles (``Parametric''). 

Using the parameterized inflow profiles rather than the mean TNBL velocity profiles in the synthetic inflow method has a small effect on the wake recovery. 
Across all Rossby numbers, the wake recovery, which we quantify with the streamtube-averaged velocity deficit in \cref{fig:wakes_linear_tnbl}($a$), agrees well between all three inflows. 
Furthermore, the qualitative trends in wake deflection are also captured when the parameterized inflow profiles are used in the synthetic inflow environment. 
The critical Rossby number corresponding to zero wake deflections is also quantitatively predicted by the simple, parameterized inflow. 
However, at low Rossby numbers, the quantitative wake deflection is underpredicted.
The underprediction in wake deflection at low Rossby numbers is due to an increase in the lateral pressure forcing in the momentum budget, particularly near the rotor (not shown). 
Nonetheless, the qualitative trend and, most importantly, the bifurcation point between positive and negative deflections are captured by the simple, parameterized inflows, which are used throughout \cref{sec:results_exploration}. 
Future work may investigate the effects on wake evolution for other inflow profile shapes (power law, low-level jet).

\end{appendix}

\bibliography{references.bib}

\end{document}